# Time-dependent nucleation rate measurements in $BaO \cdot 2SiO_2$ and $5BaO \cdot 8SiO_2$ glasses

Xinsheng Xia[a], D.C. Van Hoesen[b], Matthew E. McKenzie[c], Randall E. Youngman[c], Ozgur Gulbiten[c], K. F. Kelton[a,b]

[a] Institute of Materials Science and Engineering, Washington University, St. Louis, MO 63130, USA
[b] Department of Physics, Washington University, St. Louis, MO 63130, USA
[c] Science and Technology Division, Corning Incorporated, Corning, NY 14831, USA

Corresponding author: Xinsheng Xia, x.xia@wustl.edu

## Abstract

The two-step heat treatment method is used to measure the steady state crystal nucleation rate and induction time as a function of temperature in $BaO \cdot 2SiO_2$ and $5BaO \cdot 8SiO_2$ glasses. For both glasses, the temperature for maximum nucleation rate and the temperature range for significant nucleation agree well with previous estimates from differential thermal analysis studies. The data are analyzed with a new iterative method to obtain the interfacial free energy and critical work of cluster formation as a function of temperature. For temperatures below the temperature at which steady-state nucleation rate is a maximum, the critical work of cluster formation is shown to strongly deviate from expectations of the classical theory of nucleation.

## 1. Introduction

Understanding and controlling crystal nucleation is critically important for the manufacturing of glass and glass-ceramic products [1-3]. Nucleation and growth of crystal phases must be effectively suppressed for glass formation and subsequent processing, as well as to ensure long-term stability of the glass products. For glass ceramics, nucleation and growth must be precisely controlled to produce the desired devitrified microstructures, including the number, type, and size of crystallites formed in the glass ceramics [4-7]. Currently, the knowledge of how to control nucleation is largely gained from empirical studies. Improved models beyond the commonly used Classical Theory of Nucleation (CNT), which is known to be significantly flawed [8], are needed. The development of these more advanced models, however, is partially hindered by the lack of experimental data over a wide range of silicate glasses.

In this publication, we present new time-dependent nucleation data in two barium silicate glasses. The data were obtained using a two-step heat-treatment method [9-10], in which glasses are first heated isothermally at a temperature that is near the peak nucleation rate (the nucleation step). The nuclei formed there do not grow significantly since the growth rates are very small at this temperature. To grow them to observable size, the nucleated glasses are given a heat treatment at a second, higher, temperature where the growth rate is large, but the nucleation rate is small (the growth step).

While the nucleation rate has been measured in $BaO{\cdot}2SiO_2$ glasses in several previous studies [11-13], those results suffer from a lack of sufficient data to accurately obtain the steady state nucleation rate and one case reports inconsistent results. Further, no measurements of the nucleation rate in $5BaO{\cdot}8SiO_2$ glasses exist. In this work, the steady state nucleation rate and the induction time are measured as a function of temperature in $BaO{\cdot}2SiO_2$ and $5BaO{\cdot}8SiO_2$ glasses. The results are in agreement with previous estimates of the temperature ranges for significant nucleation and the peak temperatures in the steady-state rates obtained from differential thermal analysis (DTA) studies [14], further validating the DTA method to make quick surveys of nucleation temperatures. The experimental data are analyzed using a new iterative method to obtain the interfacial free energy and the critical work of cluster formation (nucleation barrier) as a function of temperature. As found in other silicate glasses, the interfacial free energy has a positive temperature dependence at temperatures above that of the peak nucleation rate. Below the peak nucleation temperature, the rates are different than predicted from CNT. The departure is more pronounced in these glasses than in other glasses that have been studied. They suggest that rather than continuing to decrease with decreasing temperature, the nucleation barrier first enters a plateau and then begins to rise slowly below the peak nucleation rate temperature.

## 2. Experiments

The $BaO{\cdot}2SiO_2$ and $5BaO{\cdot}8SiO_2$ glasses studied were prepared by Corning Incorporated. To produce the glasses, 2500 g batches of barium carbonate and high-purity silica were mixed in specific ratios, corresponding to the composition of each glass, and melted in platinum crucibles at 1600 °C for 6h. These liquids were then quenched to form glasses that were broken into glass cullet. To make homogenous glasses, these were re-melted at 1500 to 1600 °C for 6h, and then roller quenched or made into patties by pouring onto a stainless steel table. Glasses were analyzed using Inductively Coupled Plasma – Optical Emission Spectroscopy (ICP-OES) to determine the actual compositions, confirming that the $BaO{\cdot}2SiO_2$ and $5BaO{\cdot}8SiO_2$ glasses contained 33.2 and 38.8 mol% BaO, respectively. The glass transition temperature ($T_g$) was determined by differential scanning calorimetry (DSC) with a heating rate of 10 K/min after cooling the samples at 10 K/min from the supercooled liquid to room temperature. The $T_g$ values for the two glasses are 694.5 °C for $BaO{\cdot}2SiO_2$ and 696.6 °C for $5BaO{\cdot}8SiO_2$.

One 5BaO·8SiO2 glass was heat treated at 725 °C for 97 minutes and then heat treated at 846 °C for 47 minutes to crystallize the sample. The crystallized sample surface was polished using 400 grit SiC paper, washed in water and ultrasonically cleaned in acetone. Using this sample, the enthalpy of fusion (212.3 kJ/mol) and the liquidus temperature (1446.4 °C) were obtained from DSC measurements using a NETZSCH DSC 404 F1 Pegasus. The temperature and heat flow sensitivity of the DSC were calibrated by using the melting point of standard metals and the phase transition points of inorganic compounds. The heat flow accuracy was further verified by using a sapphire standard. The DSC measurements were made in platinum crucibles under an inert argon atmosphere at a 10 K/min heating rate from room temperature.

The $BaO{\cdot}2SiO_2$ and $5BaO{\cdot}8SiO_2$ glasses were cut into smaller samples for the nucleation studies. The $BaO{\cdot}2SiO_2$ samples were approximately 5 mm x 5 mm in area dimension and 1.24±0.08 mm thick; the $5BaO{\cdot}8SiO_2$ had the same area dimensions and were 1.33±0.15 mm thick. For the two-step heat treatments, the samples were contained in a Coors high alumina combustion boat and placed in the center zone of a Lindberg Blue M three-zone furnace. The nucleation treatments for the $BaO{\cdot}2SiO_2$ glass samples were made at 650, 675, 700, 712, 725, 738, 750 and 775 °C, for different times. The $5BaO{\cdot}8SiO_2$ glass samples were nucleated at 675, 700, 712, 725, 738, 750 and 775 °C, also for different times. After the nucleation treatment, the $BaO{\cdot}2SiO_2$ and $5BaO{\cdot}8SiO_2$ samples were heated at 840 °C and 846 °C, respectively, to grow the nuclei to observable-sized crystals.

Following the two-step heat treatments, the samples were polished with 400, 600, and 800 grit SiC papers and a 0.5 μm $CeO_2$ suspension (Allied High Tech Products Inc.). To rule out possible surface crystallization, at least 150 μm thickness of the surfaces of the samples were removed by polishing. The samples were then etched in a 0.2 HCl 0.5 HF (vol %) water solution for 10 seconds. After etching, the samples were washed in deionized water, then ultrasonically cleaned in acetone and deionized water separately, and dried on tissue paper in the air.

Microscopy images were obtained from different surface polished regions of the partially crystallized glass using an Olympus BX41M-LED optical microscope. A typical image for a $BaO{\cdot}2SiO_2$ heat-treated glass is shown in Figure 1; the crystals have an irregular shape. For this case, the area of the crystal in the image was measured, treating the irregular shape as a sphere with a diameter corresponding to the measured area of the crystal. The correction equation for spherical crystals (eq. 1) was then used to calculate the number of crystals per unit volume, $N_V$

$$N_V = \frac{2}{\pi} N_S \bar{Y} \ , \qquad (1)$$

where $N_S$ is the number of crystals measured per area in the image and $\bar{Y}$ is the average of the reciprocal diameters in the image [15-16]. Overlapping crystals were not used to determine the diameter, but were still counted to obtain the number of crystals per unit area. If the overlapping was so great that the number of crystals could not be counted, the image was not used for the analysis. Instead, a different growth treatment time was chosen to produce less crystal overlap. A typical image of the crystals in the $5BaO{\cdot}8SiO_2$ glass is shown in Figure 2; unlike the crystals in $BaO{\cdot}2SiO_2$, these have spherical shapes. The diameters of each crystal in the image could then be measured directly and used to compute $N_V$ . This was repeated for the data obtained for all of the nucleating temperatures. The standard deviation in $N_V$ was calculated from the different images at each nucleation treatment.

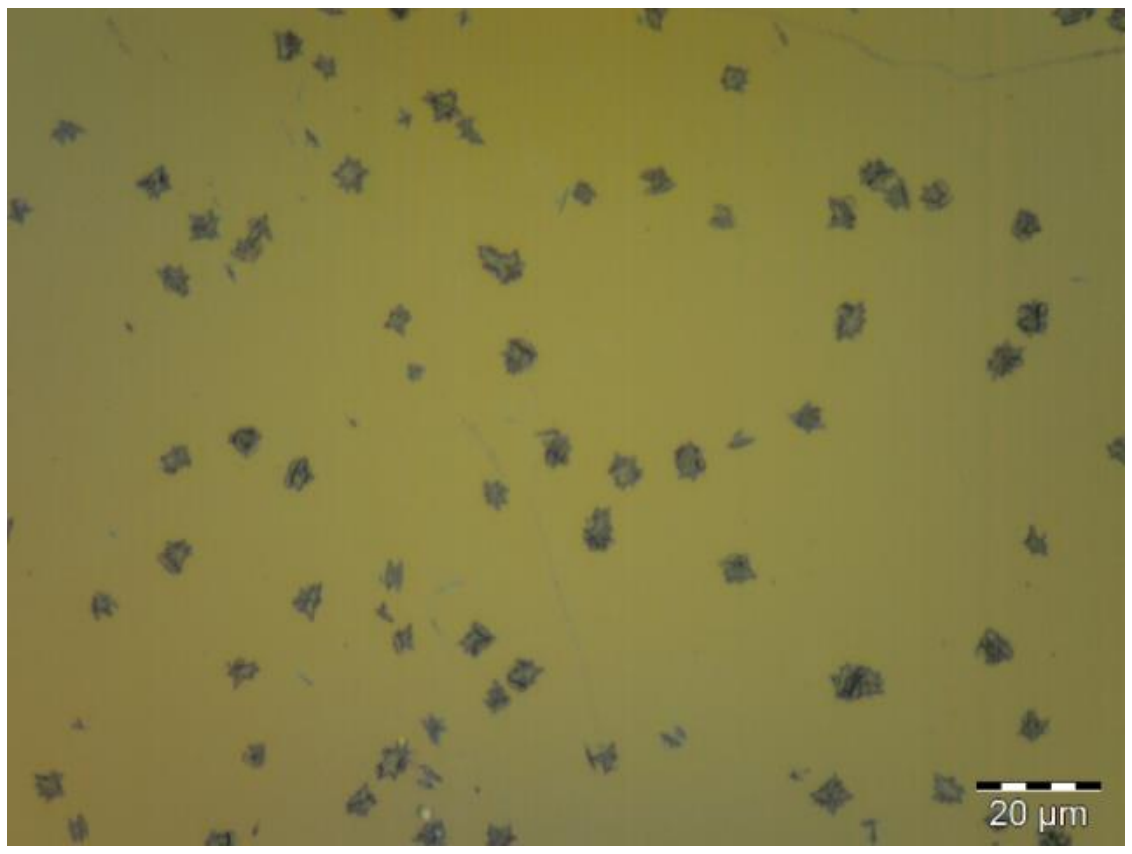

**Figure 1.** Typical image observed in optical microscope for a $BaO{\cdot}2SiO_2$ glass that was heat treated at 725 °C for 9 minutes to develop a population of nuclei; these were grown to observable size by a second heat treatment at 840 °C.

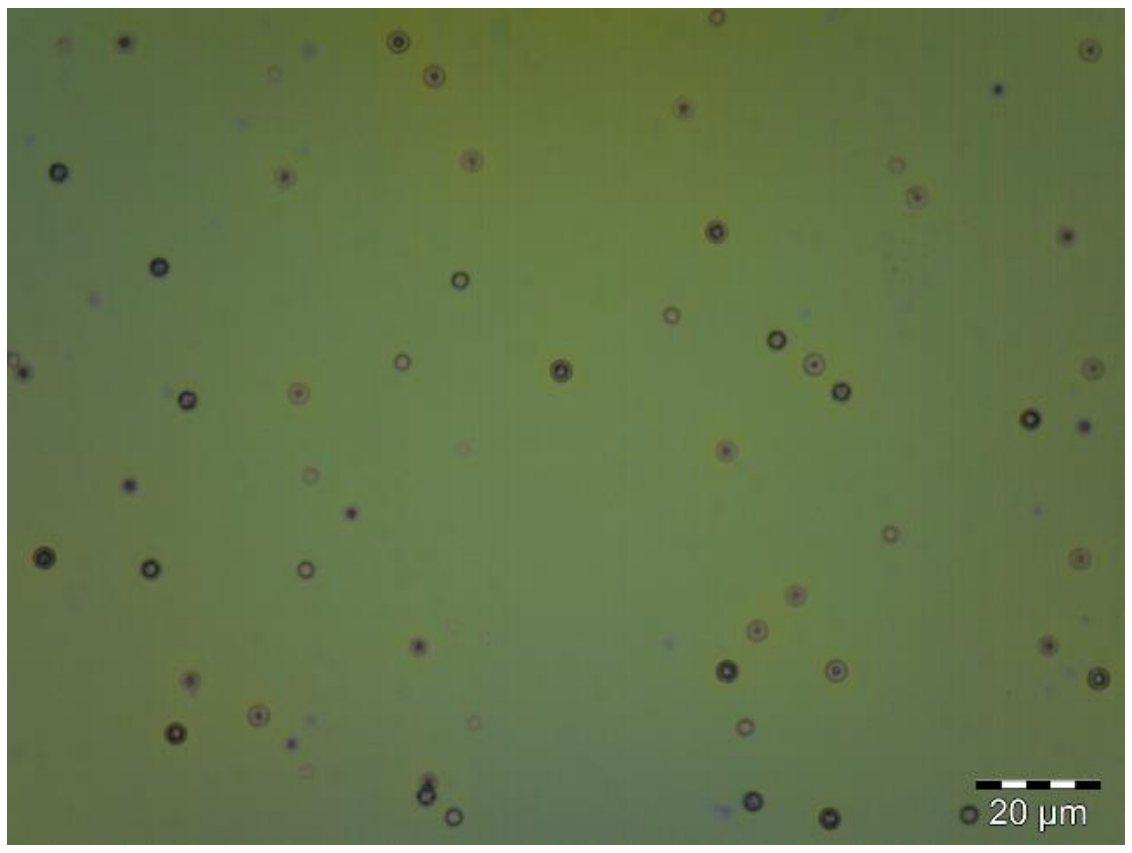

**Figure 2.** Typical image observed in optical microscope for $5BaO{\cdot}8SiO_2$ glass that was heat treated at 700 °C for 57 minutes to develop a population of nuclei; these were grown to observable size by a second heat treatment at 846 °C.

## 3. Results

### *3.1 $BaO{\cdot}2SiO_2$*

The values of $N_V$ for the $BaO{\cdot}2SiO_2$ glass are shown as a function of the heat treatment time at the nucleating temperature in Figure 3. Initially, $N_V$ increases nonlinearly with time, reflecting a time-dependent nucleation rate. After some time it increases linearly with time, reflecting a steady-state nucleation rate; the steady-state nucleation rate, $I^{st}$, is given by the slope of $N_V$ vs. time in this region. The intercept of the line with the time axis is the induction time at the critical size for the growth temperature, $\theta_{n*}^{T_G}$ [8]. The measured values for $I^{st}$ and $\theta_{n*}^{T_G}$ are listed in Table 1. The induction times are very short for the temperatures 738 °C, 750 °C and 775 °C and could not be obtained from the data; the negative intercepts are not physical but reflect the measurement uncertainty.

As shown in Figure 4, the measured steady state nucleation rate has a maximum at 712 °C. This is in agreement with a previous estimate made from DTA studies [14]. The measured width of the nucleation rate (i.e. the temperature range for significant nucleation) is also in agreement with the DTA estimates. The range is approximately 675 °C to 775 °C for the directly measured nucleation rate data presented here. The range estimated in the DTA studies was from approximately 660 °C to 770 °C [14].

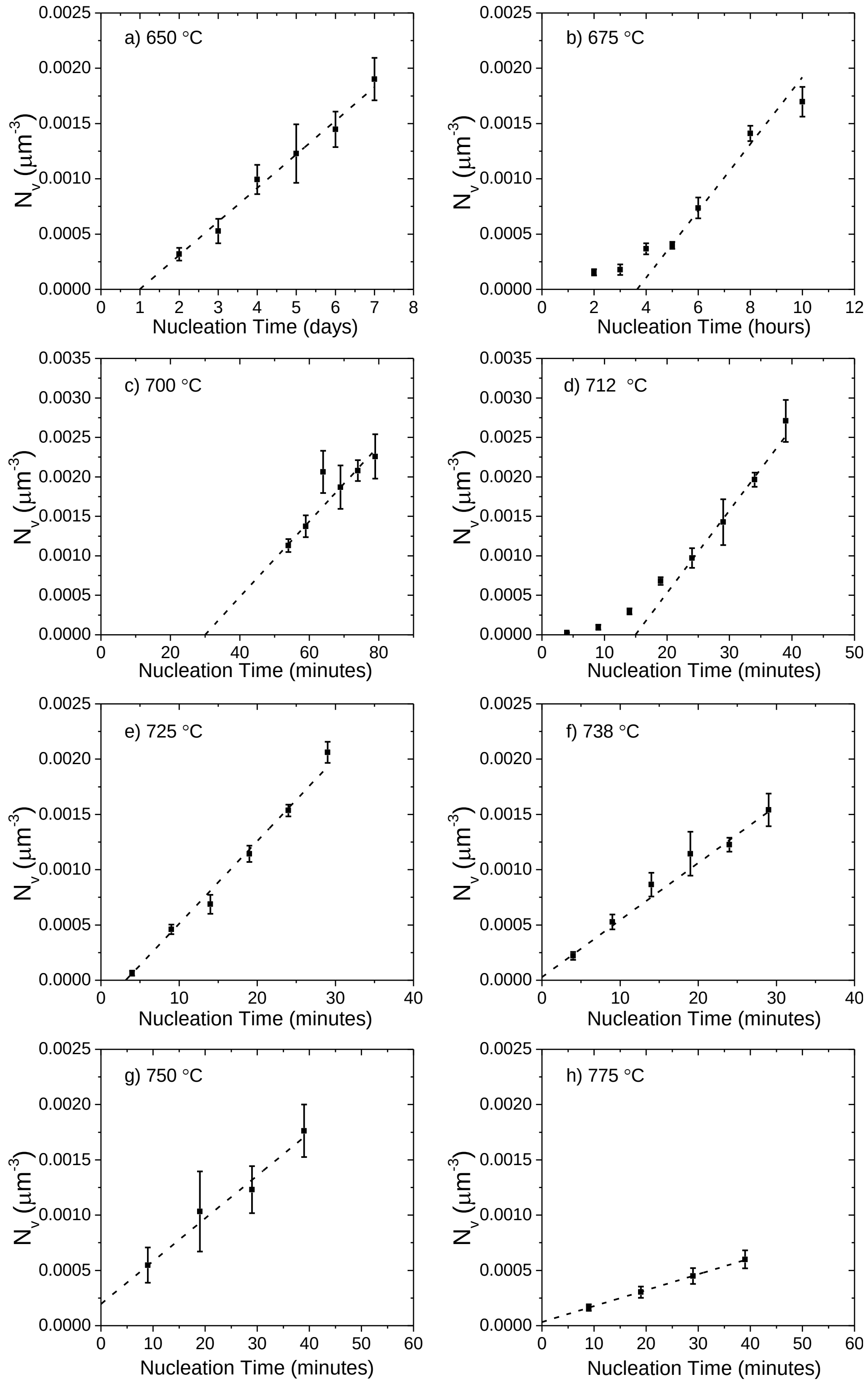

**Figure 3.** $N_V$ as a function of time at the nucleation temperatures studied for the BaO·2SiO₂ glasses. The dashed lines show the linear fits in the steady-state used to determine $I^{st}$. (The error bars on the data are equal to the standard deviation in $N_V$ calculated from the different microscope images at each temperature and heat treating time.)

**Table 1**

Steady state rates and induction times for nucleation in $BaO{\cdot}2SiO_2$ glasses

| Temperature, $T$(°C) | Steady State Nucleation Rate, $I^{st}$ ($mm^{-3}s^{-1}$) | Induction Time, $\theta_{n*}^{T_G}$ (minutes) |
|---|---|---|
| 650 | 3.5 ± 0.2 | 1423.9 ± 199.6 |
| 675 | 84.0 ± 8.4 | 218.1 ± 14.9 |
| 700 | 797.3 ± 94.4 | 30.0 ± 3.8 |
| 712 | 1745.2 ± 134.1 | 14.9 ± 1.3 |
| 725 | 1246.3 ± 41.2 | 3.2 ± 0.3 |
| 738 | 860.9 ± 41.5 | Not Determined |
| 750 | 644.9 ± 57.5 | Not Determined |
| 775 | 239.7 ± 2.9 | Not Determined |

* Mean and standard deviation values are obtained from the linear fit in the $N_V$ vs. nucleation time graphs using instrumental weighting in Origin software.

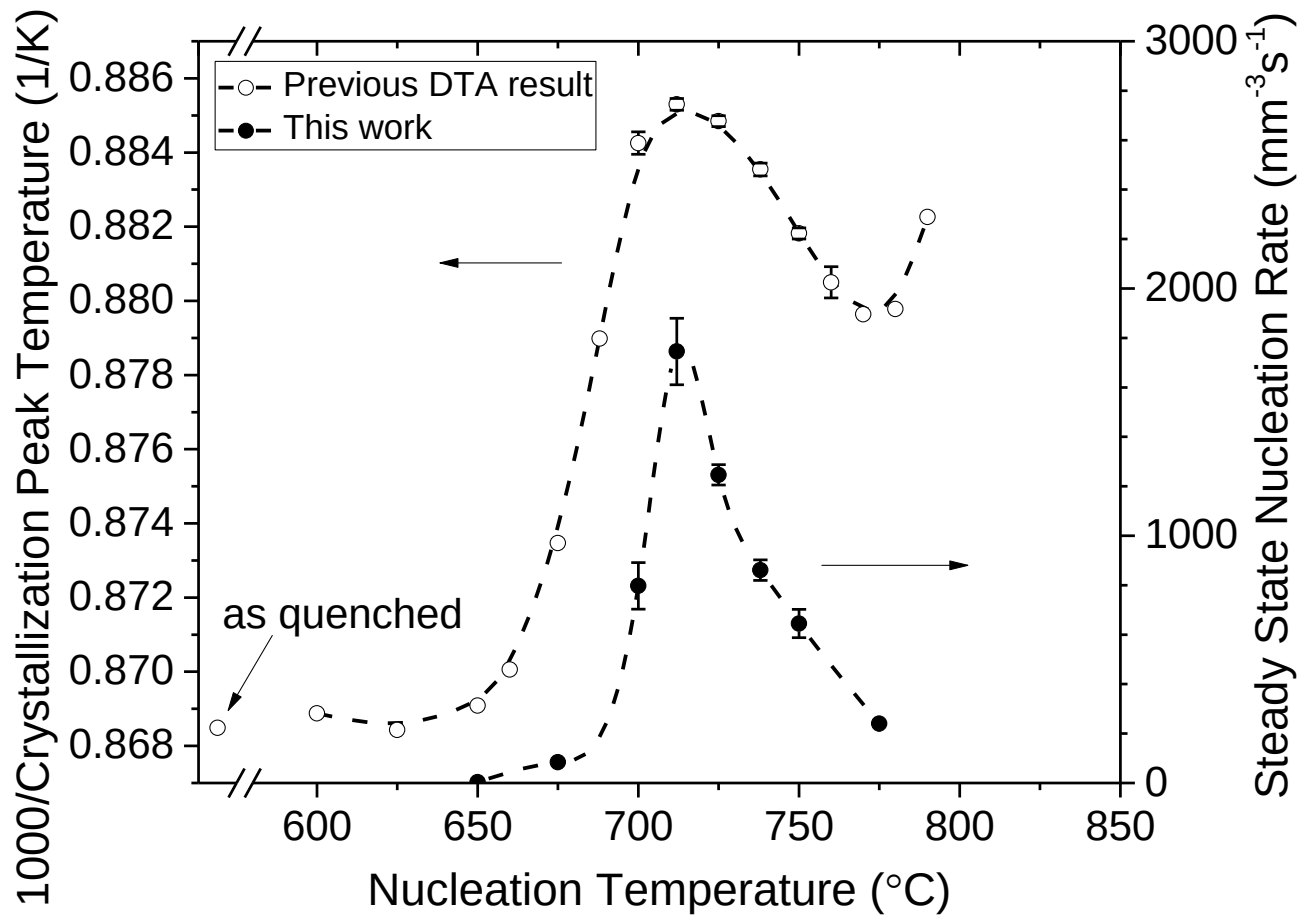

**Figure 4.** The steady state nucleation rate and the inverse DTA crystallization peak temperature (which tracks with the nucleation rate - see [14]) versus the nucleation temperature for $BaO{\cdot}2SiO_2$ glasses.

### *3.2 $5BaO{\cdot}8SiO_2$*

The measured values of $N_V$ for the $5BaO{\cdot}8SiO_2$ glass are shown as a function of the heat treatment time at the nucleating temperature in Fig. 5. As for the $BaO{\cdot}2SiO_2$ glass, the crystal nucleation rate is time-dependent, reaching the steady-state value after a sufficiently long heat treatment at the nucleation temperature. The values of $I^{st}$ and $\theta_{n*}^{T_G}$, obtained from a linear fit to the steady-state region, are summarized in Table 2. The induction time at 775 °C could not be obtained since the intercept was negative, again reflecting the measurement error for short induction times.

The measured steady state nucleation rate has a maximum at 725 °C (Figure 6). Like the $BaO{\cdot}2SiO_2$ glass, this result is in agreement with an estimation made from previous DTA studies [14]. The temperature ranges for significant nucleation in the data measured here are also similar to those estimated in the DTA studies.

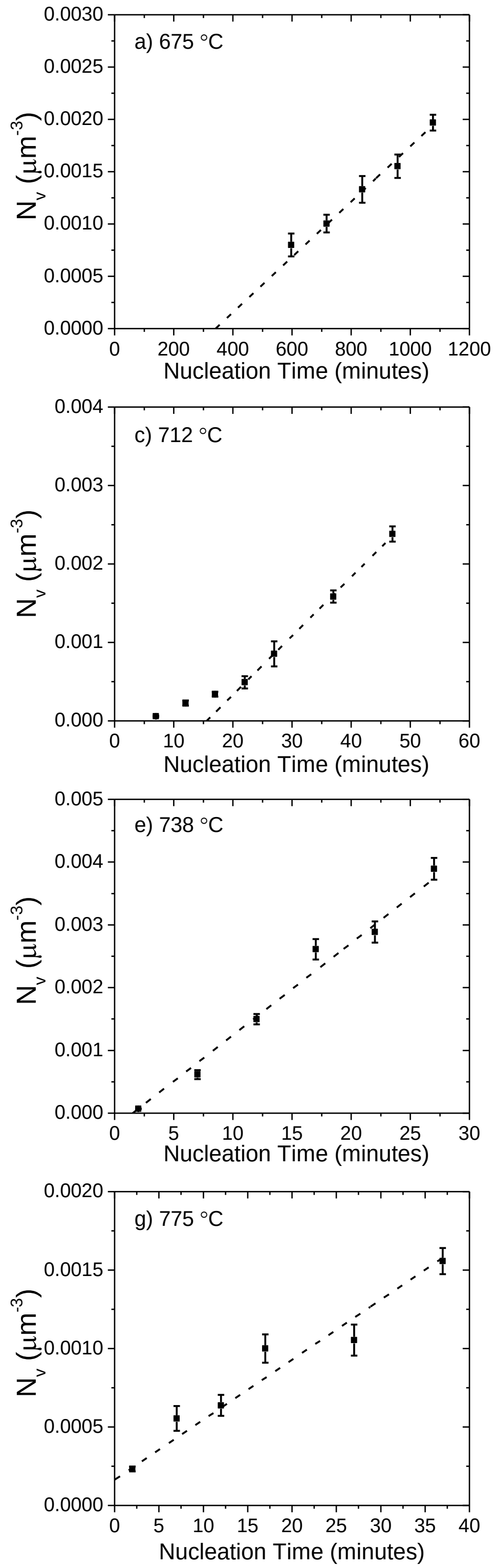

**Figure 5.** $N_V$ as a function of time at the nucleation temperatures studied for the $5BaO{\cdot}8SiO_2$ glasses. The dashed lines show the linear fits in the steady-state used to determine $I^{st}$. (The error bars on the data are equal to the standard deviation in $N_V$ calculated from the different microscope images at each temperature and heat treating time.)

**Table 2**
Steady state rates and induction times for nucleation in $5BaO{\cdot}8SiO_2$ glasses

| **Temperature, $T$(°C)** | **Steady State Nucleation Rate, $I^{st}$ (mm$^{-3}$s$^{-1}$)** | **Induction Time, $\theta_{n*}^{T_G}$ (minutes)** |
|---|---|---|
| 675 | 44.1 ± 2.8 | 341.2 ± 37.7 |
| 700 | 702.8 ± 69.3 | 44.4 ± 4.3 |
| 712 | 1251.3 ± 22.3 | 15.6 ± 0.4 |
| 725 | 3006.9 ± 51.1 | 7.3 ± 0.2 |
| 738 | 2446.3 ± 120.3 | 1.5 ± 0.1 |
| 750 | 1926.0 ± 26.7 | 0.8 ± 0.1 |
| 775 | 637.3 ± 48.9 | Not Determined |

* Mean and standard deviation values are obtained from the linear fit in the $N_V$ vs. nucleation time graphs using instrumental weighting in Origin software.

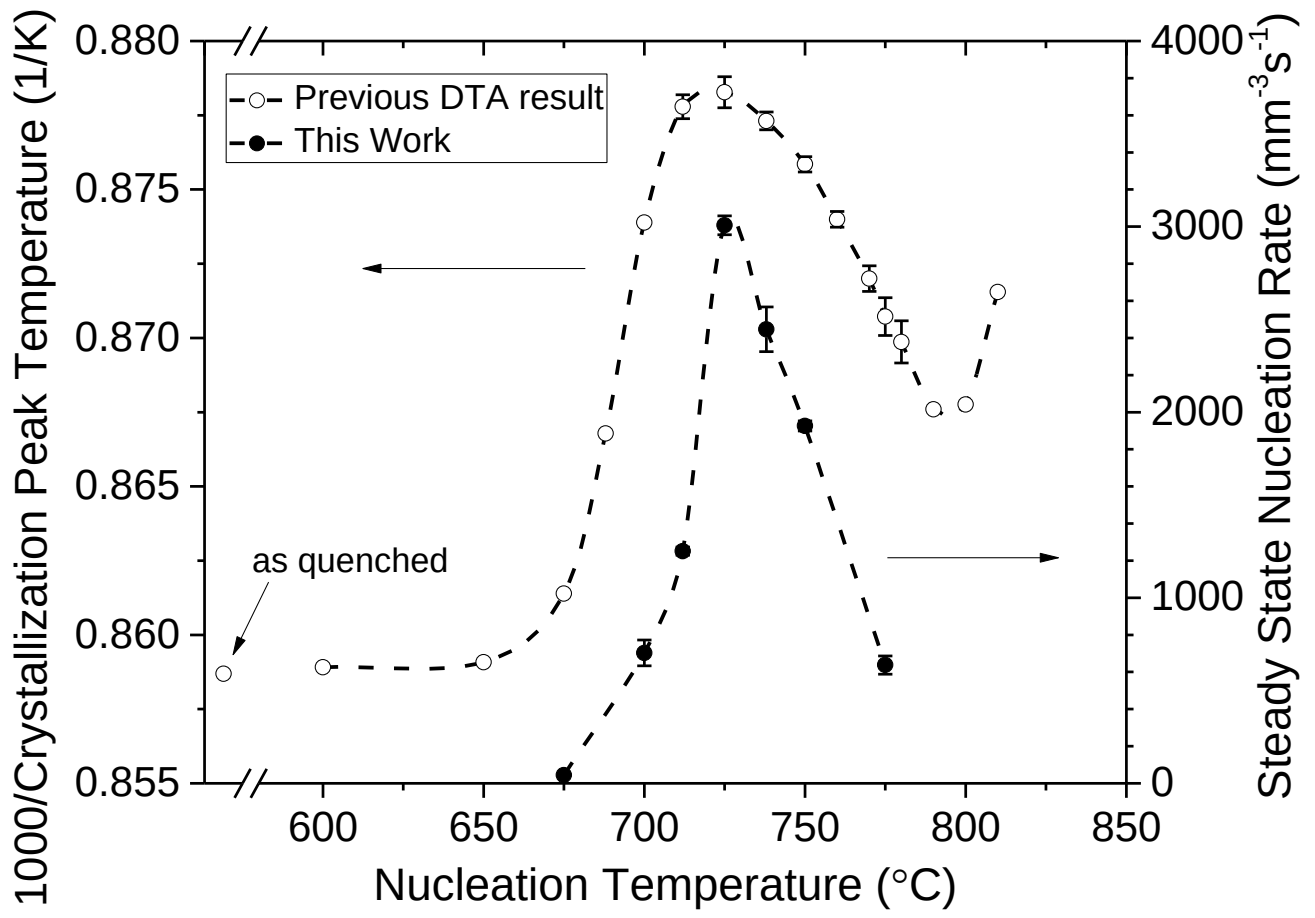

**Figure 6.** Steady state nucleation rate and inverse DTA crystallization peak temperature [14] versus nucleation temperature for $5BaO{\cdot}8SiO_2$ glasses.

## 4. Methods of Analysis

### *4.1 Method #1*

The measured data are analyzed using the CNT [8] for homogeneous nucleation of spherical clusters, with $I^{st}$ (number per volume per second) given by

$$I^{st} = \frac{k_{n*}^{+} Z N_A}{V_m} \exp\left(-\frac{W^*}{k_B T}\right). \qquad (2)$$

Here $k_{n*}^{+}$ is the rate of single monomer attachment to the critical cluster of size $n^*$, $Z$ is the Zeldovich factor, $N_A$ is the Avogadro's number, $V_m$ is the molar volume, $W^*$ is the critical work of cluster formation, $k_B$ is the Boltzmann constant and $T$ is the temperature in absolute units. For spherical clusters, $n^*$ is given by

$$n^* = \frac{32\pi}{3\bar{\mathrm{v}}} \frac{\sigma^3}{|\Delta g_{\mathrm{v}}|^3} , \tag{3}$$

where $\sigma$ is the interfacial free energy, $\bar{\mathrm{v}}$ is the volume of a monomer $\left(\bar{\mathrm{v}} = V_m / N_A\right)$, and $\Delta g_{\mathrm{v}}$ is the Gibbs free energy difference between the initial and nucleating phase per unit volume. For spherical clusters, the critical work of cluster formation, $W^*$, is

$$W^* = \frac{16\pi}{3} \frac{\sigma^3}{|\Delta g_{\mathrm{v}}|^2} . \tag{4}$$

The Zeldovich factor in eq. (2) is given by

$$Z = \left( \frac{|\Delta\mu|}{6\pi k_B T n^*} \right)^{1/2} , \tag{5}$$

where $\Delta\mu$ is the change in chemical potential for a monomer moving from the initial phase to the nucleating phase. $\Delta\mu$ is related to $\Delta g_{\mathrm{v}}$ as $\Delta\mu = \Delta g_{\mathrm{v}} \bar{\mathrm{v}}$. From the Kashchiev treatment [17-18], the induction time at the critical size for the nucleating temperature, $\theta$, is related to the transient time, $\tau_K$

$$\theta = \frac{\pi^2}{6} \tau_K , \tag{6}$$

which is a fundamental time describing the evolution of the cluster distribution and is given by

$$\tau_{\mathrm{K}} = -\frac{24 k_{\mathrm{B}} T n^*}{\pi^2 k_{n^*}^+ \Delta\mu} = \frac{4}{\pi^3 k_{n^*}^+ Z^2} . \tag{7}$$

Since the forward rate constant, $k_{n^*}^+$ can be extracted from the induction time, there is no need to assume a relation between $k_{n^*}^+$ and the diffusion coefficient or viscosity in the parent phase. The product of the steady state nucleation rate and the induction time, then, contains no kinetic terms,

$$I^{st}\theta = \frac{2}{3\pi \bar{\mathrm{v}}\, Z}\exp\left(-\frac{W^*}{k_B T}\right). \tag{8}$$

Taking the natural logarithm of eq. (8) and using the values for $W^*$ in eq. (4) and $Z$ in eq. (5),

$$\ln\left(I^{st}\theta\right) = \ln\left(\frac{16}{3\bar{\mathrm{v}}^2}\right) + \frac{1}{2}\ln\left(\frac{\sigma^3 k_B T}{\left|\Delta g_{\mathrm{v}}\right|^4}\right) - \left(\frac{16\pi}{3k_B}\frac{1}{T\left|\Delta g_{\mathrm{v}}\right|^2}\right)\sigma^3. \tag{9}$$

The induction time measured in the two-step experiments is the value at the critical size for the growth temperature, $T_G$, i.e. $\theta_{n*}^{T_G}$. However, $\theta$ in eq. (9) is the induction time at the critical size for the nucleation temperature, $T_N$, i.e. $\theta_{n*}^{T_N}$. The time required for a cluster to grow from $n^*$ at $T_N$ to $n^*$ at $T_G$ must therefore be taken into account. An equation that relates $\theta_{n*}^{T_N}$ to $\theta_{n*}^{T_G}$ is [8,19],

$$\frac{\theta_{n*}^{T_G}}{\theta_{n*}^{T_N}} = \frac{6}{\pi^2}\left[\xi + \ln\xi + \ln\left(\frac{6W^*}{k_{\mathrm{B}}T}\right) + \zeta_{\mathrm{E}} - 1\right], \tag{10}$$

where $\zeta_{\mathrm{E}}$ is Euler's constant (0.5772…) and

$$\xi = \left(\frac{n_{T_G}^*}{n_{T_N}^*}\right)^{1/3} - 1 = \left(\frac{\sigma_{T_G}}{\sigma_{T_N}}\right) \Bigg/ \left(\frac{\left|\Delta g_{\mathrm{v}}\right|_{T_G}}{\left|\Delta g_{\mathrm{v}}\right|_{T_N}}\right) - 1\ ; \tag{11}$$

$W^*$ is calculated from eq. (4), using $\sigma_{T_N}$ and $\left|g_{\mathrm{v}}\right|_{T_N}$. By using eq. (10) and eq. (11), eq. (9) can be modified to have the form

$$\begin{aligned}\ln\left(I^{st}\theta_{n^*}^{T_G}\right) = \ln\Bigg\{&\left(\frac{\sigma_{T_G}}{\sigma_{T_N}}\right)\Bigg/\left(\frac{\left|\Delta g_{\mathrm{v}}\right|_{T_G}}{\left|\Delta g_{\mathrm{v}}\right|_{T_N}}\right) + \ln\left[\left(\frac{\sigma_{T_G}}{\sigma_{T_N}}\right)\Bigg/\left(\frac{\left|\Delta g_{\mathrm{v}}\right|_{T_G}}{\left|\Delta g_{\mathrm{v}}\right|_{T_N}}\right) - 1\right] + \ln\left[\frac{32\pi}{k_B}\right] + \ln\left[\frac{\left(\sigma_{T_N}\right)^3}{T_N\left(\left|\Delta g_{\mathrm{v}}\right|_{T_N}\right)^2}\right] + \zeta_{\mathrm{E}} - 2\Bigg\} \\ &+ \ln\left[\frac{32}{\pi^2\bar{\mathrm{v}}^2}\right] + \frac{1}{2}\ln\left(\frac{\left(\sigma_{T_N}\right)^3 k_B T_N}{\left(\left|\Delta g_{\mathrm{v}}\right|_{T_N}\right)^4}\right) - \left(\frac{16\pi}{3k_B}\frac{1}{T_N\left(\left|\Delta g_{\mathrm{v}}\right|_{T_N}\right)^2}\right)\left(\sigma_{T_N}\right)^3 .\end{aligned} \tag{12}$$

$I^{st}$ and $\theta_{n^*}^{T_G}$ were obtained from the data presented here, while $\bar{\mathrm{v}}$ and $\Delta g_{\mathrm{v}}$ at $T_N$ and $T_G$ were obtained from the literature or from experimental measurements. One monomer is assumed to be one formula unit. The only two unknown parameters in eq. (12) are, therefore, the interfacial free energies at the growth and nucleation temperatures, $\sigma_{T_G}$ and $\sigma_{T_N}$ respectively. An iterative method is followed to determine these: (a) assume an initial value for $\sigma_{T_G}$ ; (b) use the measured values of $I^{st}$ and $\theta_{n^*}^{T_G}$ at each nucleating temperature to calculate $\sigma_{T_N}$ from eq. (12); (c) for $T_N$'s at the temperature of the maximum steady state nucleation rate and above, linearly extrapolate $\sigma_{T_N}$ to the growth temperature to obtain a new value for $\sigma_{T_G}$ ; (d) if the difference between the initial value of $\sigma_{T_G}$ and the new one is larger than $10^{-5}$ J/m$^2$, take the new value of $\sigma_{T_G}$ and perform the calculation again until it converges; (e) using the converged value of $\sigma_{T_G}$ , calculate $\sigma_{T_N}$ from eq. (12) for $T_N$ with known $I^{st}$ and $\theta_{n*}^{T_G}$ ; (f) linear extrapolate $\sigma_{T_N}$ of temperatures at and above the temperature for the maximum nucleation rate to higher temperatures to get $\sigma_{T_N}$ for $T_N$ whose $I^{st}$ is known but $\theta_{n*}^{T_G}$ is unknown; (g) calculate $W^*/k_BT$ and $\theta_{n*}^{T_N}$ for different $T_N$ with eqs. (4) and (10).

*4.2 Method #2*

It is widely assumed that the interfacial free energy obtained from nucleation data linearly increases with increasing temperature [8]. This is argued to be a consequence of the diffuse interface between the nucleating cluster and the parent phase [20-23] . Using $\sigma_{T_N}$ for $T_N$ at the temperature of the maximum steady state nucleation rate and above obtained from method #1, a linear fit with temperature was therefore made to get new $\sigma_{T_N}$ for $T_N$ lower than the temperature of maximum steady state nucleation rate. These new values for $\sigma_{T_N}$ were used to calculate new $\left|\Delta g_{\mathrm{v}}\right|_{T_N}$ for these low temperatures from eq. (12). This was particularly important for the low nucleating temperatures where the critical sizes are small and the values computed from the thermodynamic properties measured for large samples might not be appropriate to describe small clusters. Finally, $W^* / k_B T$ and $\theta_{n^*}^{T_N}$ for these low temperatures were calculated from eqs. (4) and (10).

## 5. Analysis Results and Discussion

From eq. (9), a graph of $\ln\left(I^{st}\theta_{n^*}^{T_N}\right)$ as a function of $\dfrac{1}{T_N\left(\left|\Delta g_v\right|_{T_N}\right)^2}$ (where $T_N$ is the nucleating temperature) should produce a straight line. Using eq. (10) to obtain $\theta_{n^*}^{T_N}$ from $\theta_{n^*}^{T_G}$, this graph is shown in fig. 7. For fig. 7, the $\Delta g_v$ used was from the Turnbull approximation, which is the same as later discussed in sections 5.1.1 and 5.2.1; a constant value for $\sigma$ was assumed (0.100 J/m$^2$) [17].

While a straight line is observed at higher temperatures in both glasses, a marked departure near the peak nucleation temperature is observed. While this has been observed previously in other silicate glasses [24], it is particularly strong in these measurements. The reasons for this behavior remain unclear. This is explored in more detail in the following sections.

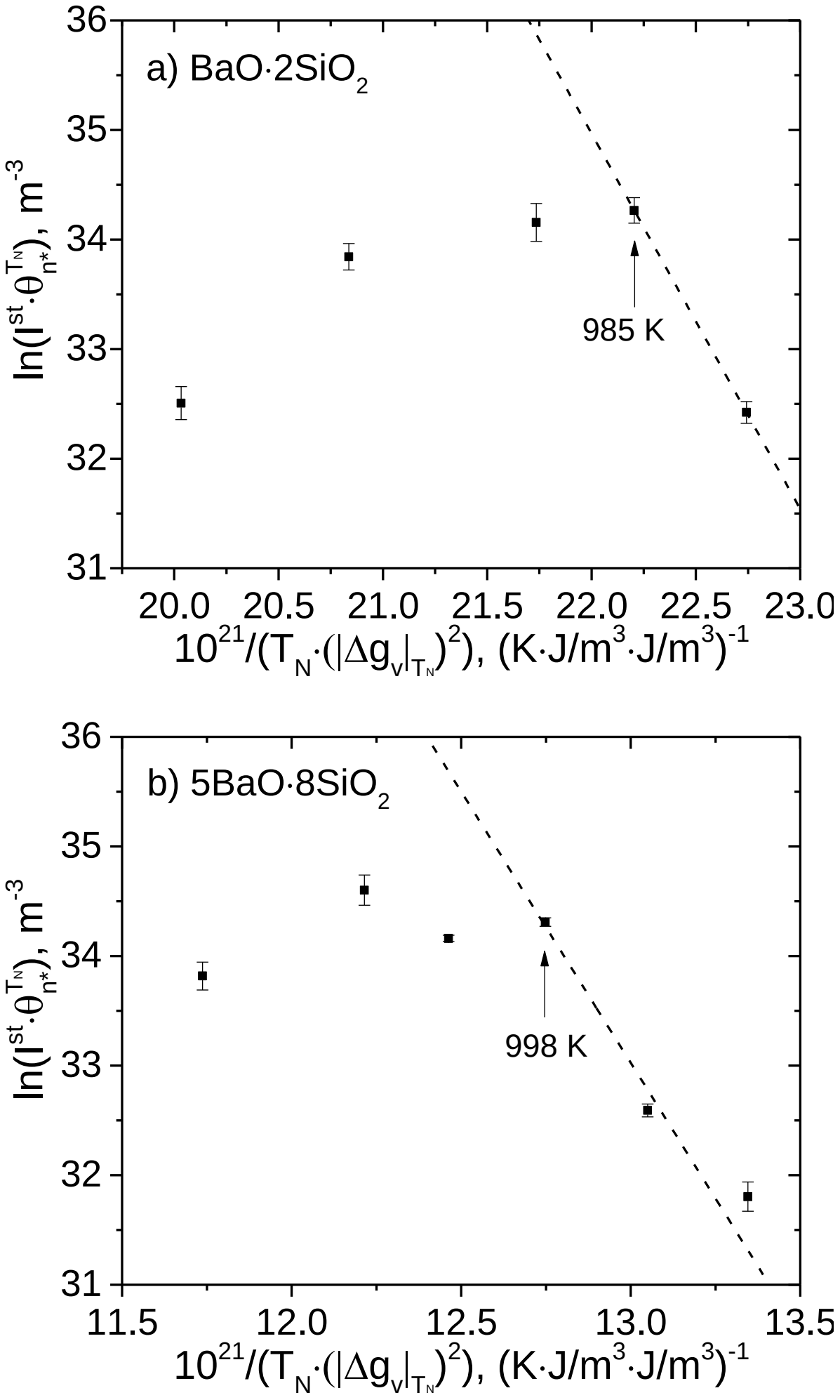

**Figure 7.** $\ln\left(I^{st}\theta_{n*}^{T_N}\right)$ as a function of $\dfrac{1}{T_N\left(\left|\Delta g_{\mathrm{v}}\right|_{T_N}\right)^2}$ in (a) $BaO{\cdot}2SiO_2$ and (b) $5BaO{\cdot}8SiO_2$ glasses showing a linear behavior at high temperatures (the dashed lines).

*5.1 BaO·2SiO2*

*5.1.1 Results from Method #1*

For the analysis of the data for the $BaO{\cdot}2SiO_2$ glasses, one monomer was assumed as one formula unit of monoclinic $BaO{\cdot}2SiO_2$ and $V_m$ was $73.34 \times 10^{-6}\ \mathrm{m^3/mol}$ [25]. The Turnbull approximation was used to calculate $\Delta g_{\mathrm{v}}$ with an enthalpy of fusion equal to $37.5\ \mathrm{kJ/mol}$ and a liquidus temperature of 1420 °C [26]; the temperature-dependent values for $\left|\Delta g_V\right|$ are shown in

Figure 8a. They increase with decreasing temperature, as expected. The initial value of $\sigma_{T_G}$ was chosen to be $0.106\ \mathrm{J/m^2}$; the converged value was $0.118\ \mathrm{J/m^2}$. The values computed for $\sigma_{T_N}$, $W^*/k_BT$ and $\theta_{n^*}^{T_N}$ from eqs. (12), (4), and (10) using the converged $\sigma_{T_G}$ are shown as a function of the nucleation measurement temperature in Figure 8. The solid points in Fig. 8.b are the values for $\sigma_{T_N}$ obtained by this procedure. The unfilled squares are the values of $\sigma$ that result from a linear extrapolation of the values at and above the peak nucleation temperature to higher temperatures. The unfilled points in Fig. 8.c for $W^*/k_BT$ are calculated using the extrapolated points in Fig. 8.b for $\sigma$.

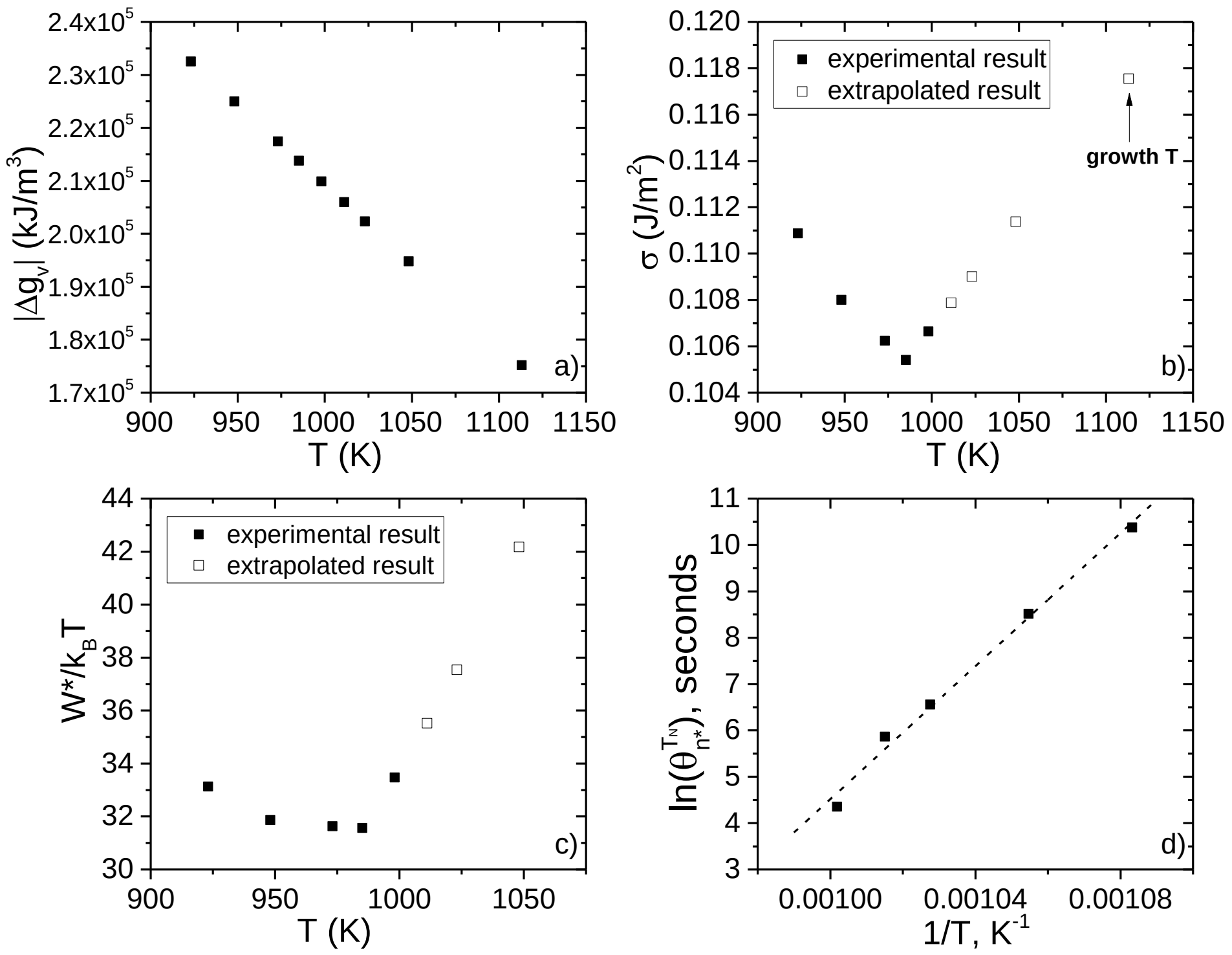

**Figure 8.** The values of $\sigma_{T_N}$, $W^*/k_BT$ and $\theta_{n^*}^{T_N}$ from the measured nucleation data for the $BaO\cdot 2SiO_2$ glasses: (a) $|\Delta g_V|$ as a function of temperature, (b) the calculated temperature dependence of the interfacial free energy, (c) the calculated values for $W^*/k_BT$ as a function of temperature, (d) the log of the induction time as a function of inverse temperature.

Radically different results are observed for temperatures above or below the temperature for the maximum steady-state nucleation rate (985 K). As shown in Fig. 8.b, the interfacial free energy increases above 985 K, as has commonly been found. However, it decreases with increasing temperature below 985 K. As shown in Fig. 8.c, $W^*/k_BT$ decreases with decreasing temperature at high temperatures, as expected within the CNT [27], but it plateaus and begins to slowly increase with decreasing temperature below 985 K. As normally observed, the log of the corrected induction time, $\theta_{n*}^{T_N}$ , increases with increasing inverse temperature (Figure 8.d). From the Kashchiev expression eq. (7), this reflects the increased mobility at higher temperatures.

### *5.1.2 Results from Method #2*

Why the interfacial free energy would have such an abrupt change in temperature dependence below the peak nucleation rate temperature is difficult to understand. Based on other evidence, it should increase linearly with temperature [20-22]. To investigate this, the high temperature data in Fig. 8.b were linearly extrapolated to lower temperature to force this agreement and to investigate further the deviation from CNT at lower temperatures (923K, 948K and 973K). The result of the extrapolation is shown in Figure 9a. Using these values for $\sigma_{T_N}$ , $\left|\Delta g_v\right|_{T_N}$ was calculated from eq. (12). As shown in Figure 9b, this assumption of a linear temperature dependence causes $\left|\Delta g_V\right|$ to depart from the values expected from the Turnbull approximation and begin to decrease with decreasing temperature at 985 K. A similar behavior to that discussed in Fig. 8.c is again found for $W^*/k_BT$ (Fig. 9.c), with the value plateauing and then increasing with decreasing temperature, which is not the behavior expected from CNT, or even more advanced theories such as the diffuse interface theory or the semi-empirical density functional theory [20-22,28]. The temperature dependence of $\theta_{n*}^{T_N}$ remains essentially unchanged (Fig. 9.d).

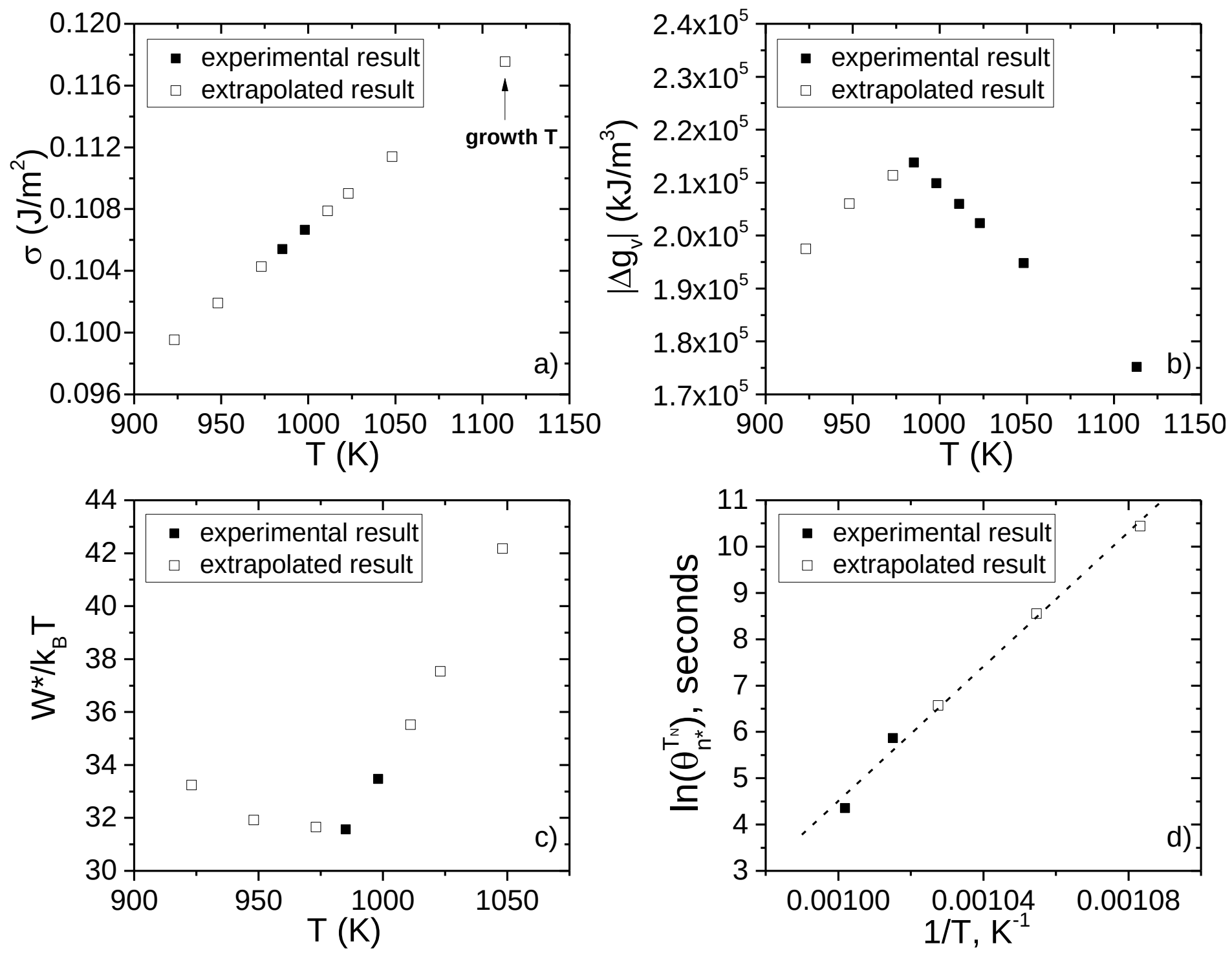

**Figure 9.** The values of $W^*/k_BT$ and $\theta_{n^*}^{T_N}$ from the measured nucleation data for the $BaO{\cdot}2SiO_2$ glasses, assuming a linear temperature dependence of the interfacial free energy, $\sigma_{T_N}$ (a): (b) the calculated values of $|\Delta g_V|$ as a function of temperature; (c) the calculated values for $W^*/k_BT$ as a function of temperature; (d) the log of the induction time as a function of inverse temperature.

### *5.2 5BaO·8SiO₂*

#### *5.2.1 Results from Method #1*

The nucleation data in the $5BaO{\cdot}8SiO_2$ glass were analyzed following method similar discussed in section 4.1. One monomer was assumed as one formula unit of $5BaO{\cdot}8SiO_2$ and the value for $V_m$ was taken as $317.7 \times 10^{-6}$ m³/mol, which was calculated from the monoclinic $5BaO{\cdot}8SiO_2$ structure (ICSD 100311). The value for $\Delta g_v$ was calculated from the Turnbull approximation using the values of the enthalpy of fusion (212.3 kJ/mol) and the liquidus temperature (1446.4 °C), both obtained from DSC measurements. The values of $\Delta g_v$ as a function of temperature are shown in Fig. 10.a. The initial value of $\sigma_{T_G}$ was 0.113 J/m²; it converged to 0.131 J/m$^2$. The calculated values for $\sigma_{T_N}$, $W^*/k_BT$ and $\theta_{n*}^{T_N}$ using the converged value of $\sigma_{T_G}$ are shown in Fig. 10.b – 10.d.

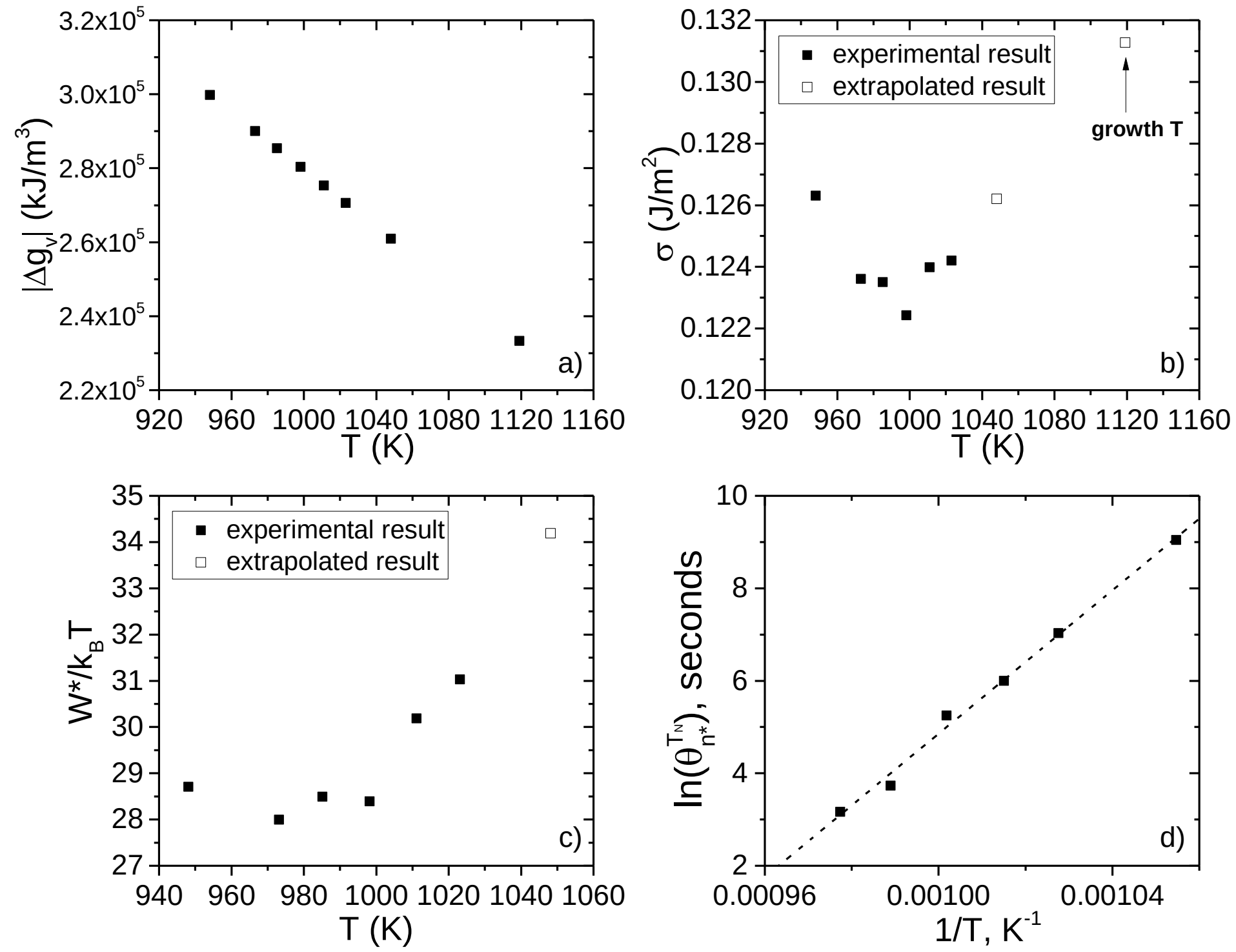

**Figure 10.** The values of $\sigma_{T_N}$, $W^*/k_BT$ and $\theta_{n*}^{T_N}$ from the measured nucleation data for the 5BaO·8SiO$_2$ glasses: (a) $|\Delta g_V|$ as a function of temperature, (b) the calculated temperature dependence of the interfacial free energy, (c) the calculated values for $W^*/k_BT$ as a function of temperature, (d) the log of the induction time as a function of inverse temperature.

As for the BaO·2SiO$_2$ glass, the calculated values of $\sigma_{T_N}$ in the 5BaO·8SiO$_2$ glass change the sign of their temperature dependence near the peak nucleation temperature (998 K). Again, $W^*/k_BT$ enters a plateau below this temperature (Fig. 10.c), in disagreement with CNT and other more advanced nucleation theories. Also, as for BaO·2SiO$_2$ glass, the logarithm of the induction time increases linearly with inverse temperature, reflecting the Arrhenius temperature dependence of the atomic mobility.

*5.2.2 Results from Method #2*

Again following the approach already discussed for the $BaO{\cdot}2SiO_2$ glass, the high temperature values of $\sigma$ were linearly extrapolated to lower temperatures (Fig. 11.a) and the values of $\left|\Delta g_{\mathrm{v}}\right|_{T_N}$, $W^*/k_BT$ and $\theta_{n*}^{T_N}$ were calculated using these values of $\sigma$. Again, this causes $\left|\Delta g_V\right|$ to decrease below the peak nucleation rate temperature (Figure 11.b). The values of $W^*/k_BT$ still reach a plateau below that temperature (Fig. 11.c), which is in conflict with all know theories of nucleation. The logarithm of the induction time still scales with inverse temperature (Fig. 11.d).

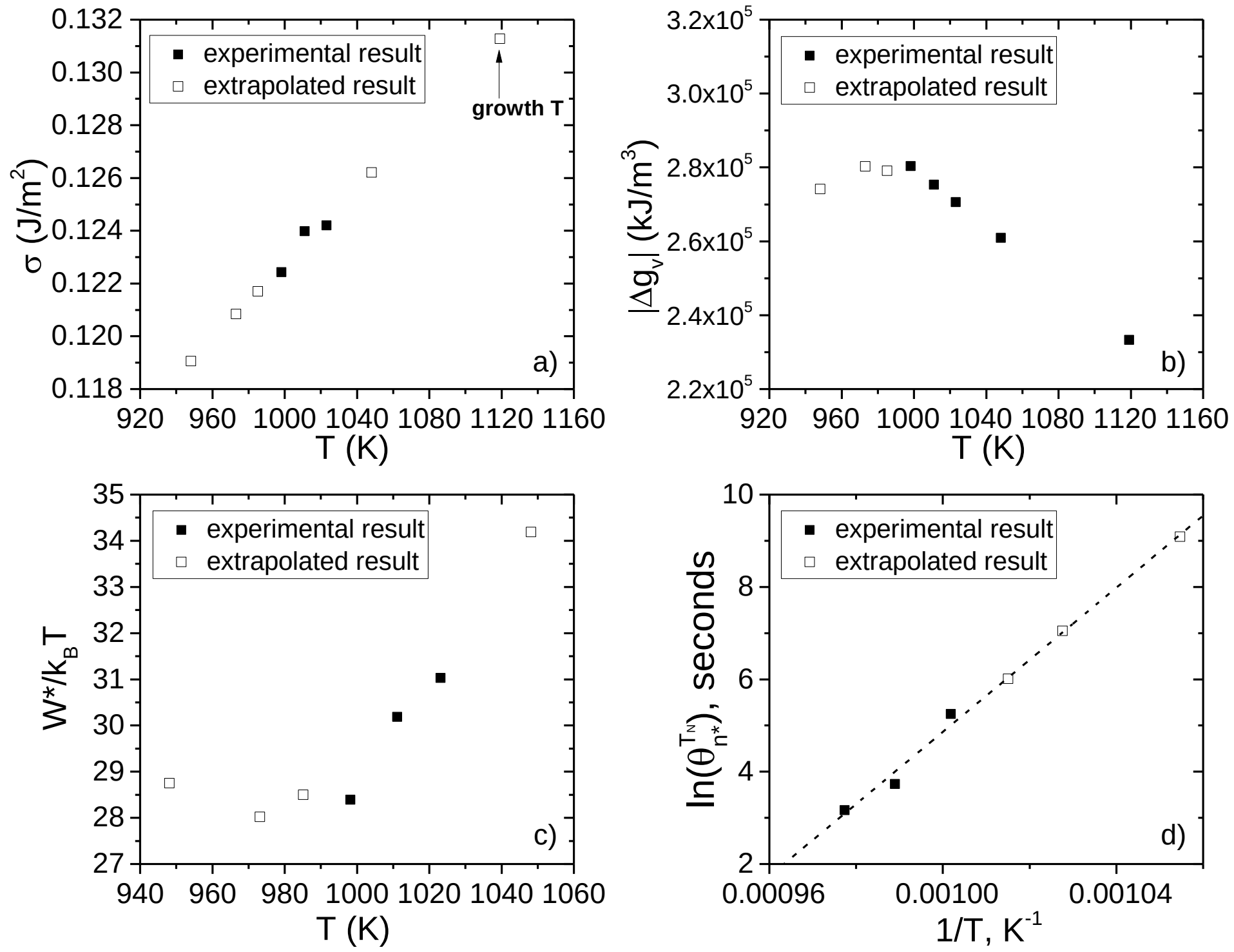

**Figure 11.** The values of $W^*/k_BT$ and $\theta_{n*}^{T_N}$ from the measured nucleation data for the $5BaO{\cdot}8SiO_2$ glasses, assuming a linear temperature dependence of the interfacial free energy, $\sigma_{T_N}$ (a): (b) the calculated values of $\left|\Delta g_V\right|$ as a function of temperature; (c) the calculated values for $W^*/k_BT$ as a function of temperature; (d) the log of the induction time as a function of inverse temperature.

A consistent picture emerges from these studies. In both glasses, there is a radical change of trend in the critical work of cluster formation (nucleation barrier) for temperatures below the peak in the nucleation rate. A similar behavior has been observed previously in other silicate glasses [27,29]. While sometimes explained as experimental error, the effect is far outside experimental error in these $mBaO\cdot nSiO_2$ studies. Some investigators have tested possible theoretical explanations such as elastic stress, changing size of structural units, spatial heterogeneity and dynamical heterogeneity to account for the effect [27,29-31], but the correct explanation remains unclear due to the lack of experimental or modeling evidence.

## 6. Conclusions

In summary, the steady state nucleation rates and induction times were measured at multiple temperatures in $BaO\cdot 2SiO_2$ and $5BaO\cdot 8SiO_2$ glasses using the two-step heat treatment method. For $BaO\cdot 2SiO_2$ glass, the steady-state nucleation rate has a maximum at 712 °C. For $5BaO\cdot 8SiO_2$ glass the maximum rate is at 725 °C. These results are in agreement with previous estimates made from differential thermal analysis measurements.

A new iterative method was introduced to calculate the interfacial free energy and critical work of cluster formation. For both glasses, the critical work of cluster formation shows an anomalous behavior, entering a plateau and slowly increasing rather than decreasing with decreasing temperature at low temperatures. This behavior either arises from an abrupt change in the temperature dependence of the interfacial free energy or decrease in the driving free energy for nucleation. The change in driving free energy is more likely, since the positive temperature dependence of the interfacial free energy arises from the diffuse interface between the cluster and the parent phase, which probably becomes even more diffuse at lower temperatures. The departure in the temperature dependence of the nucleation barrier is in disagreement with all known theories of nucleation. As we will explain in a future publication, using a Grand Canonical Monte Carlo simulation technique [32], we suggest that it arises from a closing-off of the nucleation pathways at lower temperatures. Clearly, nucleation processes in silicate glasses are much more complicated than has been understood previously.

## Acknowledgements

The authors thank L. Cai, A. Gangopadhyay, R. Ashcraft, R. Dai, M. Sellers, R. Chang for useful discussions. This work is financially supported by a NSF GOALI grant, DMR 17-20296, and by Corning Incorporated.

## References

[1] J.C. Mauro, C.S. Philip, D.J. Vaughn, M.S. Pambianchi, Glass science in the United States: current status and future directions, Int. J. Appl. Glass Sci. 5 (2014) 2-15.

[2] J.C. Mauro, E.D. Zanotto, Two centuries of glass research: historical trends, current status, and grand challenges for the future, Int. J. Appl. Glass Sci. 5 (2014) 313-327.

[3] J.C. Mauro, Grand challenges in glass science, Frontiers in Materials 1 (2014) 20.

[4] W. Höland, V. Rheinberger, M. Schweiger, Control of nucleation in glass ceramics, Philos. Trans. Royal Soc. A 361 (2003) 575-589.

[5] W. Holand, G.H. Beall, Glass ceramic technology, John Wiley & Sons, 2012.

[6] M.J. Davis, E.D. Zanotto, Glass-ceramics and realization of the unobtainable: Property combinations that push the envelope, MRS Bulletin 42 (2017) 195-199.

[7] G.H. Beall, Design and properties of glass-ceramics, Annu. Rev. Mater. Sci. 22 (1992) 91-119.

[8] K.F. Kelton, A.L. Greer, Nucleation in Condensed Matter-Applications in Materials and Biology, Elsevier, Amsterdam, 2010

[9] P.F. James, Kinetics of crystal nucleation in lithium silicate-glasses, Phys. Chem. Glasses 15 (1974) 95-105.

[10] K.L. Narayan, K.F. Kelton, First measurements of time-dependent nucleation as a function of composition in $Na_2O \cdot 2CaO \cdot 3SiO_2$ glasses, J. Non-Cryst. Solids 220 (1997) 222-230.

[11] E.G. Rowlands, PhD Thesis, University of Sheffield, 1976.

[12] E.D. Zanotto, P.F. James, Experimental tests of the classical nucleation theory for glasses, J. Non-Cryst. Solids 74 (1985) 373-394.

[13] A.M. Rodrigues, PhD Thesis, Federal University of São Carlos, 2014.

[14] X. Xia, I. Dutta, J.C. Mauro, B.G. Aitken, K.F. Kelton, Temperature dependence of crystal nucleation in $BaO \cdot 2SiO_2$ and $5BaO \cdot 8SiO_2$ glasses using differential thermal analysis, J. Non-Cryst. Solids 459 (2017) 45-50.

[15] R.T. DeHoff, F.N. Rhines, Determination of number of particles per unit volume from measurements made on random plane sections: the general cylinder and the ellipsoid, Trans. AIME 221 (1961) 975-982.

[16] E.D. Zanotto, PhD Thesis, University of Sheffield, 1982.

[17] K.F. Kelton, Crystal Nucleation in Liquids and Glasses, Solid State Physics, Vol. 45, Academic Press, New York, 1991.

[18] D. Kashchiev, Solution of the non-steady state problem in nucleation kinetics, Surf. Sci. 14 (1969) 209-220.

[19] V.A. Shneidman, M.C. Weinberg, Induction time in transient nucleation theory, J. Chem. Phys. 97 (1992) 3621–3628.

[20] F. Spaepen, Homogeneous nucleation and the temperature dependence of the crystal-melt interfacial tension, Solid State Physics, Vol. 47, Academic Press, Boston, 1994.
[21] L. Gránásy, Diffuse interface theory of nucleation, J. Non-Cryst. Solids 162 (1993) 301-303.
[22] L. Gránásy, Diffuse interface model of crystal nucleation, J. Non-Cryst. Solids 219 (1997) 49-56.
[23] D. Turnbull, Kinetics of solidification of supercooled liquid mercury droplets, J. Chem. Phys. 20, (1952) 411-424.
[24] M.C. Weinberg, E.D. Zanotto, Re-examination of the temperature dependence of the classical nucleation rate: homogeneous crystal nucleation in glass, J. Non-Cryst. Solids 108 (1989) 99-108.
[25] G. Oehlschlegel, Binary partial $BaO{\cdot}2SiO_2$-$2BaO{\cdot}3SiO_2$ system, Glastech. Ber. 44 (1971) 194-204.
[26] E.G. Rowlands, P.F. James, The Structure of Non-Crystalline Solids, edited by P. H. Gaskell, p. 215, Taylor and Francis, London, 1977.
[27] A.S. Abyzov, V.M. Fokin, A.M. Rodrigues, E.D. Zanotto, J.W. Schmelzer, The effect of elastic stresses on the thermodynamic barrier for crystal nucleation, J. Non-Cryst. Solids, 432 (2016) 325-333.
[28] C.K. Bagdassarian, D.W. Oxtoby, Crystal nucleation and growth from the undercooled liquid: A nonclassical piecewise parabolic free-energy model, J. Chem. Phys. 100 (1994) 2139-2148.
[29] V.M. Fokin, A.S. Abyzov, E.D. Zanotto, D.R. Cassar, A.M. Rodrigues, J.W.P. Schmelzer, Crystal nucleation in glass-forming liquids: Variation of the size of the “structural units” with temperature, J. Non-Cryst. Solids 447 (2016) 35-44.
[30] A.S. Abyzov, V.M. Fokin, N.S. Yuritsyn, A.M. Rodrigues, and J.W.P. Schmelzer, The effect of heterogeneous structure of glass-forming liquids on crystal nucleation, J. Non-Cryst. Solids 462 (2017) 32-40.
[31] P.K. Gupta, D.R. Cassar, E.D. Zanotto, Role of dynamic heterogeneities in crystal nucleation kinetics in an oxide supercooled liquid, J. Chem. Phys. 145 (2016) 211920.
[32] M.E. McKenzie, J.C. Mauro, Hybrid Monte Carlo technique for modeling of crystal nucleation and application to lithium disilicate glass-ceramics, Comput. Mater. Sci. 149 (2018) 202-207.